\documentstyle[aps,eqsecnum,12pt]{revtex}

\begin{document}


\title{\bf Black hole polarization and new entropy bounds}

\author{Jacob D. Bekenstein\thanks{Electronic mail: bekenste@vms.huji.ac.il}
and Avraham E. Mayo\thanks{Electronic mail:
Mayo@venus.fiz.huji.ac.il}}

\address{\it The Racah Institute of Physics, Hebrew University
of Jerusalem,\\ Givat Ram, Jerusalem 91904, Israel}

\date{\today}

\maketitle

\begin{abstract}
Zaslavskii has suggested how to tighten Bekenstein's bound on
entropy when the object is electrically charged.  Recently Hod has
provided a second tighter version of the bound applicable when the
object is rotating.  Here we derive Zaslavskii's optimized  bound
by considering the accretion of an ordinary charged object by a
black hole.  The force originating from the polarization of the
black hole by a nearby charge is central to the derivation of the
bound from the generalized second law.  We also conjecture an
entropy bound for charged rotating objects, a synthesis of
Zaslavskii's and Hod's.  On the basis of the no hair principle for
black holes, we show that this last bound cannot be tightened
further in a generic way by knowledge of ``global'' conserved
charges, {\it e.g.,\/} baryon number, which may be borne by the
object.
\end{abstract}

\section{Introduction}

A universal bound on the entropy of a macroscopic object of
maximal radius $R$ bearing  energy $E$ has been proposed by one of
us\cite{Bound}:
\begin{equation}
S\leq 2\pi ER/\hbar \label{firstbound}
\end{equation}
(units with $G=c=1$ are used). This bound was first inferred by
considering the infall of the relevant object into a black hole,
and arranging for the infall conditions to cause a minimum of
horizon area growth.  Appealing to the generalized second law
(GSL)\cite{GSL} then gave bound (\ref{firstbound}) as a condition
for the overall entropy not to decrease.  This derivation was
criticized\cite{UW1} for leaving out the effects of buoyancy in
the acceleration (Unruh) radiation.  In some scenarios this makes
a difference in the energy that is added to the black hole by the
infall, and thus to its horizon area increase.  However, it has
become clear\cite{Newbound,PW} that if the Unruh acceleration
buoyancy comes from a small number of particle species (as it must
in our universe), then for objects which are not too thin in one
of their dimensions, and whose parts are described by quantum
mechanics, the buoyancy correction is indeed negligible, and one
can derive bound (\ref{firstbound}) by appealing to the GSL.

Independent support exists for bound (\ref{firstbound}).  It is
satisfied trivially for composites of nonrelativistic particles by
virtue of the fact that entropy of a system is never far from the
number of particles involved. And  for free massless quantum
fields enclosed in volumes of various shapes, the bound's validity
has been  checked directly.  Both numerical
verification\cite{Numer} and analytical proof\cite{BekSchiff}
exist (see review by Bekenstein and Schiffer\cite{Review}).  The
entropy bound can also be inferred directly from the properties of
the acceleration radiation\cite{Zaslavskii1}.

Regarding self--gravitating systems, Sorkin, Wald and
Jiu\cite{SWJ} gave evidence that bound (\ref{firstbound}) is valid
for thermal radiation on the verge of gravitational collapse,
while Zaslavskii\cite{Zaslavskii2} proves the bound for a system
consisting of a static black hole in equilibrium with thermal
radiation in a box.  When the black hole has charge $e$,
Zaslavskii\cite{Zaslavskii3} infers the tighter bound
\begin{equation}
S\leq {2\pi ER\over \hbar}\left(1-{e^2\over 2ER}\right),
\label{zaslavskii}
\end{equation}
although he admits to some uncertainty regarding the numerical
coefficients. Zaslavskii does not claim this form of the bound for
systems {\it not\/} containing a black hole.

In its original form (\ref{firstbound}), the entropy bound is
saturated by the Schwarzschild black hole [whose entropy is $4\pi
(2M)^2/4\hbar= 2\pi M\cdot (2M)/\hbar$].  This prompted the
observation\cite{Bound} that the Schwarzschild black hole is the
most entropic object for given size and energy.  But the Kerr
black hole's entropy falls below bound (\ref{firstbound}) (this
will be true for any reasonable interpretation of the radius $R$
of the nonspherical Kerr hole).

This asymmetric state of affairs motivated Hod\cite{Hod} to search
for a tighter bound on entropy for objects with angular momentum
which is saturated by the Kerr hole.  Hod repeats Bekenstein's
derivation\cite{BHentropy,Brazil} of the minimal increment in
Kerr-Newman (KN) horizon area that is caused by an object's
infall.  That derivation applied the idea of
Christodoulou\cite{Christodoulou} together with
Carter's\cite{Carter,MTW} integrals of the Lorentz equations of
motion to a particle of rest mass $\mu$ and radius $R$ moving in a
KN background.   The minimal growth in horizon area was found from
the conservation laws and the relation they establish between the
change in black hole parameters and the energy and orbital angular
momentum of a particle in an orbit such that the particle's center
of mass (CM) can get to distance $R$ from the horizon:
\begin{equation}
(\Delta A)_{\rm min} = 8\pi\mu R \label{damin}
\end{equation}
It is remarkable that this minimal area growth is independent of
the black hole parameters.  Because $\mu$ can be identified with
the total proper energy of the object, bound (\ref{firstbound})
follows from (\ref{damin}) and the GSL.

Spin of the  particle was not taken into account in Carter's
integrals.  Hod refers instead to Hojman and Hojman's\cite{Hojman}
integrals of motion for a neutral object with spin $s$ moving on a
KN background.  Repeating the argument that led to
Eq.(\ref{damin}), Hod gets
\begin{equation}
(\Delta A)_{\rm min} = 8\pi\mu R (1-s^2\mu^{-2} R^{-2})^{1/2}
\label{Hod1}
\end{equation}
Appeal to the GSL then allows Hod to infer
\begin{equation}
S\leq {2\pi ER\over \hbar}\left(1-{s^2\over E^2R^2}\right)^{1/2}
\label{Hod2}
\end{equation}
As Hod remarks, a Kerr black hole of mass $E=m$ and spin $s=j\leq
m$ exactly saturates bound (\ref{Hod2}) provided one identifies
$R$ with $(r_+^2+j^2/m^2)^{1/2}$, where
$r_+=m+(m^2-j^2/m^2)^{1/2}$ is the radial Boyer--Lindquist
coordinate for the Kerr horizon.  The identification is reasonable
because $4\pi(r_+^2+j^2/m^2)$ is exactly the area of the Kerr
horizon.

In Sec.II we take up the question of how to derive bound
(\ref{zaslavskii}) for an ordinary charged object (not a system
including a black hole as in Zaslavskii\cite{Zaslavskii3}) by
analogy with the original derivation of bound (\ref{firstbound})
using the GSL.  In Sec.III we calculate the change in horizon area
that results from lowering a charged object into an electrically
isolated black hole, and thus furnish a derivation from the GSL of
Zaslavskii's bound (\ref{zaslavskii}).  Sec.IV contains a variant
using an electrically grounded black hole; it leads to the same
result.  In Sec.V we conjecture an entropy bound for rotating
charged objects, assemble supporting evidence, and also give a
partial proof that it cannot generically be made tighter by taking
other conserved quantities, {\it e.g.\/} baryon number, into
account.

\section{The role of black hole polarization}

Granted the validity of the original entropy bound
(\ref{firstbound}) for a macroscopic object, is Hod's bound
reasonable from a mundane point of view ? For small $s$
(nonrelativistic rotation) we may expand the r.h.s. of
Eq.(\ref{Hod2}) to get
\begin{equation}
S\leq {2\pi R\over \hbar}\left[E-{s^2\over 2\mu R^2}\right]
+O(s^4) \label{approx}
\end{equation}
where we have replaced $E\rightarrow \mu$ (rest mass) in the
denominator. Now an object with moment of inertia $I$ has internal
energy $\epsilon=E-s^2/2I$.  For a thin spherical shell $I$
reaches its maximum, ${\scriptstyle 2\over\scriptstyle 3}\mu R^2$,
so that the internal energy for given $E$ and $R$ is also
maximized: $\epsilon_{\rm max}=E-3s^2/(4\mu R^2)$. The phase space
available to the object's degrees of freedom is controlled by
$\epsilon$.  Hence,  if bound (\ref{firstbound}) is valid at
$s=0$, we would infer $S\leq (2\pi R/\hbar)[E-3s^2/(4\mu R^2)]$
when $s\neq 0$.  Hod's bound is a bit more liberal; as a result,
it manages to encompass the Kerr black hole.

Now for a nonrotating  object of mass $\mu$, radius $R$ and charge
$e$, the Coulomb energy attains its minimum, $e^2/2R$, when the
charge is uniformly spread on a thin shell of radius $R$.  Thus
the internal energy of the object has the maximum $\epsilon_{\rm
max}=E-{\scriptstyle 1\over\scriptstyle 2}e^2/R$.  If bound
(\ref{firstbound}) is valid at $e=0$, we expect the tighter
entropy bound $S\leq (2\pi R/\hbar) (E-{\scriptstyle
1\over\scriptstyle 2}e^2/R)$ for the charged object.  This
coincides with Zaslavskii's proposal Eq.(\ref{zaslavskii}), and
suggests its general validity.

Is Zaslavskii's proposed bound saturated by the
Reissner-Nordstr\"om (RN) black hole ?  Let the black hole's mass
be $m$ and its charge $q$.  If bound (\ref{zaslavskii}) applies to
black holes, it predicts that
\begin{equation}
S_{\rm BH} \leq {2\pi m\over
\hbar}\left(m+\sqrt{m^2-q^2}\right)\left(1-{q^2\over
2m(m+\sqrt{m^2-q^2})}\right) \label{SBH}
\end{equation}
[we set $E=m$ and $R=r_+\equiv m+\sqrt{m^2-q^2}$]. Multiplying out
the factors we see that the right hand side is precisely
$A/4\hbar$, where $A$ is the horizon area of the RN black hole,
\begin{equation}
A = 4\pi r_+^2=4\pi (m + \sqrt{m^2-q^2}\ )^2 \label{area}
\end{equation}
Thus the RN black hole saturates Zaslavskii's bound; this is a
further point in favor of the bound's validity and efficiency.  In
Sec.V we shall arrive at a hybrid bound which embodies fully the
requirement that the black hole be the most entropic state for a
given quantity of energy, charge and angular momentum.

Nobody has thus far given a derivation of bound (\ref{zaslavskii})
for charged objects patterned after those originally used to
derive bounds (\ref{firstbound}) and (\ref{Hod2}) from the GSL.
Both those derivations focused on accretion of the relevant object
by a black hole, and on the concomitant change in horizon area.
Extension of this type of argument to the charged object is {\it
not\/} straightforward.  Suppose we work with a Schwarzschild
black hole and a charged particle devoid of spin.  Naively the
particle's track is a geodesic, and so the minimal change in area
will still be given by Eq.(\ref{damin}).  In fact, if the black
hole is a KN one, the same result is obtained by using Carter's
integrals of motion for orbits of the Lorentz equation of
motion\cite{Brazil}.   Thus no improved entropy bound results for
a nonrotating charged object.  This is disturbing from the point
of view of the derivation of entropy bounds by use of the GSL: if
this approach is tenable, it should be possible to derive the
physically reasonable bound (\ref{zaslavskii}) from the GSL once
one accepts bound (\ref{firstbound}).

As we make clear in Sec.III, the mentioned problem may be traced
to the neglect of a certain force that acts on the object.  A
charged particle in a black hole's vicinity is acted upon by not
only the Lorentz force from the black hole's electromagnetic
field, but also by the (Abraham--Lorentz--Dirac) radiation
reaction force, as well as by the force originating from the black
hole's polarization by the particle's electric field.  Now it is
known that a particle at rest in a static black hole background
does not radiate (despite its being accelerated).  Hence we expect
the radiation reaction force to vanish.  This suggests focusing on
the accretion by a static black hole of an object which is lowered
slowly from a large distance to the horizon.  We can then suppose
that only the gravitational, Coulomb and polarization (image)
forces act upon it.  By this approach we succeed below in deriving
bound (\ref{zaslavskii}) by use of the GSL.   Now, {\it if\/} as
is sometimes claimed, the GSL functions independently of entropy
bounds,  there should not have been reason for an idiosyncratic
effect (black hole polarization here) to supply precisely the
missing element in the derivation of the entropy bound for charged
objects from the GSL.   This, to our mind, is the main
significance of the mentioned success: a new demonstration that
the GSL provides a valid road to entropy bounds.

\section{Lowering a charged body in a black hole's field}

We use the signature $\{-,+,+,+\}$ and denote the timelike
coordinate outside the black hole, assumed to be a spherical
static one, by $x^0$.  First consider a test particle of mass
$\mu$ and charge $e$.  Its motion, were it subject only to
gravitation and electromagnetic influences, would be governed by
the lagrangian
\begin{equation}
L = -\mu\int \sqrt{-g_{\alpha\beta} \,\dot x^\alpha \,\dot
x^\beta}\ d\tau + e\int \hat A_\alpha \,\dot x^\alpha d\tau
\label{lagrangian1}
\end{equation}
where $x^\alpha(\tau)$ denotes the particle's trajectory, $\tau$
the proper time, an overdot stands for $d/d\tau$, and $\hat
A_\alpha$ means the background electromagnetic 4--potential
evaluated at the particle's spacetime position.  Recalling that
$g_{\alpha\beta}\,\dot x^\alpha\,\dot x^\beta = -1$, it follows
from the Lagrangian that the canonical momenta are
$p_\alpha=\delta L/\delta \dot x^\alpha= \mu  g_{\alpha\beta}
\,\dot x^\beta +  e \hat A_\alpha$.  The stationarity of the
envisaged background means there is a timelike Killing vector
$\xi^\alpha=\{1,0,0,0\}$.  The quantity
\begin{equation}
{\cal E}\equiv -p_\alpha \xi^\alpha = -\mu g_{0\beta}\,\dot
x^\beta -e \hat A_0 \label{energy}
\end{equation}
is easily shown to be conserved\cite{MTW}; it corresponds to the
usual notion of energy as measured at infinity. Its first term
expands to $\mu +{1\over 2} \mu (d{\bf x}/dt)^2$ in the Newtonian
limit. The second term, $-e \hat A_0$, is thus the electric
potential energy.

In our gedanken experiment the object of rest mass $\mu$ and
charge $e$, idealized as spherically symmetric, is suspended by
some means to keep it from falling freely, and is slowly lowered
radially towards the black hole.  Of course, the forces keeping it
quasistatic change its energy measured at infinity.  The idea is
to bring the object as close to the horizon as possible, and then
drop it in, inferring from the energy measured at infinity at its
last prefall position the increase in horizon area that this
causes. A complication - the Unruh--Wald buoyancy in acceleration
radiation\cite{UW1} - may cause the object to float neutrally some
distance from the horizon, thus arresting the contemplated
descent. But as mentioned in Sec.I, provided the number of
relevant particle species in nature is not large (which seems to
be true in our universe), and provided the object is composed of
parts that obey quantum mechanics, the buoyancy is negligible all
the way to very near the horizon, and makes no practical
difference to the energetics of the process (if the object is
dropped from a bit off the horizon\cite{Newbound}).

For generality we allow the black hole to carry a charge $q$; we
require that $q$ and the charge $e$ of the object be very small on
the scale of $m$, the mass of the hole.  There are now two parts
to the electromagnetic potential: one linear in $q$, suitably
named $\hat A_0^{(q)}$,  which is produced by the black hole, and
a second one, $A_0^{(e)}$, linear in $e$, whose source is the
object itself.  Because the last is a self--field, it has no hat
symbol.  As elucidated by Vilenkin\cite{Vilenkin} and corroborated
by Smith and Will\cite{Smith}, by contrast to the situation in
flat spacetime, in the presence of a Schwarzschild black hole this
self--potential makes a nontrivial contribution to the object's
energy measured at infinity.  Those calculations are quite
involved; here we recover that effect in a much simpler way.  We
may still write the energy as measured at infinity as in
Eq.(\ref{energy}); however, we must there replace $\hat A_0$ by
$\hat A_0^{(q)}+{\scriptstyle 1\over\scriptstyle 2}A_0^{(e)}$.
The factor ${\scriptstyle 1\over \scriptstyle 2}$ is familiar from
flat--spacetime electrodynamics; it takes care of the fact that
the object owes part of its energy to its own field, not to the
background one (we thank B. Linet for reminding us of this
elementary fact). Smith and Will bring out the factor
${\scriptstyle 1\over \scriptstyle 2}$ with an explicit
calculation.

Correct to $O(e)$, which we regard as the same as $O(q)$, the
metric may be taken as Schwarzschild's.  In isotropic coordinates
it is
\begin{equation}
ds^2 = - \left({1-{m\over 2\rho}\over 1+{m\over
2\rho}}\right)^2(dx^0)^2 + (1+{m\over 2\rho})^4 \left[d\rho^2 +
\rho^2 (d\theta^2 +\sin^2 \theta\, d\phi^2)\right] \label{metric}
\end{equation}
We see that the horizon resides at $\rho=m/2$.  Because the object
is nearly stationary, its 4-velocity, which we normalize to $-1$,
must have the form $\dot x^\alpha\approx
\{(-g_{00})^{-1/2},0,0,0\}$.  Substitution in Eq.(\ref{energy})
from the metric gives for the energy, when the object's CM is at
$\rho=a$ and $\theta=0$,
\begin{equation}
{\cal E} = \mu\left(1-{m\over 2a}\right)\, \left(1+{m\over
2a}\right)^{-1} -e \left(\hat A_0^{(q)}+{\scriptstyle
1\over\scriptstyle 2}A_0^{(e)}\right)_{\rho=a, \theta=0}
\label{energy2}
\end{equation}

Eq.(\ref{potential1}) of the Appendix gives $A_0(\rho,\theta)$,
the full electromagnetic 4--potential due to a stationary (or
nearly so) point charge $e$ in the background of a spherical black
hole with small charge $q$.  This expression, accurate to
$O(e^2)$, is a trivial extension of an early brilliant solution by
Copson\cite{Copson}, as modified by Linet\cite{Linet}. Its
structure shows that one can think of the potential as getting a
contribution from image charges on the black hole.
$A_0(\rho,\theta)$ naturally diverges at $\rho=a$ and $\theta=0$,
the charge's position.  Thus if we want to use it for our finite
object, we must regularize the potential before going to the limit
$\rho\rightarrow a$ and $\theta\rightarrow 0$ as required by
formula (\ref{energy2}).

The simplest procedure is as follows.  We reexpress $A_0$ in terms
of new coordinates $\{\varrho, \vartheta,\phi\}$ centered on the
charge, rather than on the black hole center, as was the case for
$\{\rho, \theta,\phi\}$, but sharing the same polar axis.   This
implies the substitutions
\begin{equation}
\rho \cos\theta \rightarrow a + \varrho \cos\vartheta
\end{equation}
\begin{equation}
\rho\rightarrow \sqrt{a^2+\varrho^2+2a\varrho\cos\vartheta}
\end{equation}
A small metric sphere of proper radius $R$ located at $\rho=a$ and
$\theta=0$ is the coordinate sphere $\varrho=(1+{m\over
2a})^{-2}R; \forall\vartheta$. Thus it makes sense to expand $A_0$
in a Laurent series in $\varrho$:
\begin{equation}
A_0=-{\left(1-{m\over 2a}\right)\over\left(1+{m\over
2a}\right)^3}{e\over\varrho} -4a{4mae +m^2q+4aq(a+m)\over
(2a+m)^4}+C(a,m, e,q)\cos\vartheta +O(\varrho) \label{A_0}
\end{equation}
where $C$ is a complicated function independent of $\varrho$.

Now the divergent term in Eq.(\ref{A_0}) corresponds to the
Coulomb potential of the charge $e$ in flat spacetime; there we
expect $A_0=-e\varrho^{-1}$ (sign because we deal with the
covariant component).  A factor $(1-{m\over 2a}) (1+{m\over
2a})^{-1}$ is required to redshift the fourth component of a
4--vector to the point in question [see Eq.(\ref{metric}].
Finally, a factor $(1+{m\over 2a})^{-2}$ is required to convert
the coordinate distance $\varrho$ in the denominator to proper
distance.    Thus when taking the limit $\rho\rightarrow a$ and
$\theta\rightarrow 0$ ($\varrho\rightarrow 0$) of $A_0$, we must
discard the first term in the r.h.s. of Eq.(\ref{A_0}) (put
another way, the energy that comes from it is absorbed into the
renormalized rest mass $\mu$\cite{Smith}).

Our spherically symmetric finite object samples all directions
about its center without discrimination.  Because the metric also
looks isotropic in coordinates $\{\varrho, \vartheta, \phi\}$, we
must thus average out the third term in the r.h.s. of
Eq.(\ref{A_0}) over all angles $\vartheta$ and $\phi$; as a result
its contribution vanishes. Terms of $O(\varrho)$ vanish as the
size of the object shrinks.  Thus the second (constant) term in
the r.h.s. of Eq.(\ref{A_0}) furnishes the entire electrostatic
contribution to ${\cal E}$.  We separate that into the parts
$A_0^{(q)}$ (black hole's) and $\hat A_0^{(e)}$ (image charges')
defined earlier.  Substituting these in Eq.(\ref{energy2}) we find
\begin{equation}
{\cal E}=\mu\left(1-{m\over 2a}\right)\, \left(1+{m\over
2a}\right)^{-1} +4ae{2mae+q(2a+m)^2\over (2a+m)^4}
+O\left({e^4\over m^3}\right) \label{potentialenergy}
\end{equation}
The case $q=0$ of this result is equivalent to results given
earlier (in Schwarzschild coordinates) by Vilenkin\cite{Vilenkin}
for $a\gg m$ and by Smith and Will\cite{Smith} for all $a$.

When the object is near the horizon, the proper distance from its
CM to the horizon is
\begin{equation}
\ell \equiv \int_{m/2}^a (g_{\rho\rho})^{1/2} d\rho \approx
4(a-{m/2}) +O[(a-m/2)^2] \label{ell}
\end{equation}
Expressing $a$ in Eq.(\ref{potentialenergy}) in terms of $\ell$ by
means of Eq.(\ref{ell}) we get
\begin{equation}
{\cal E} =\left({2\mu\ell+e^2\over
8m}\right)\left[1+O\left({\ell\over m}\right)\right] +{eq\over 2m}
+ O\left({e^4\over m^3}\right) \label{energy3}
\end{equation}
Since we are obviously considering a black hole large and massive
compared to the object's proper radius and mass, the corrections
of $O(\ell/m)$ are appropriately neglected, as are those of
$O({e^4/m^3})$ by virtue of the assumed smallness of $e$.  The
gradual approach to the horizon must stop when the proper distance
from the object's CM to the horizon reaches the object's proper
radius $R$.  Hence,
\begin{equation}
{\cal E}\geq {2\mu R+e^2+4eq\over 8m} \label{energyfinal}
\end{equation}

As mentioned, our primary concern is with changes in the horizon
area.
 Although we have used Schwarzschild's metric in the above discussion, the
true metric is closer to that of a RN solution; it is thus best to
use as a first approximation the area appropriate to the RN black
hole, namely Eq.(\ref{area}).  This formula must be corrected for
the perturbation of the metric originating in the object, which in
linear approximation should be of $O(\mu)$ and $O(e^2)$, the first
caused by the energy momentum tensor of the object's mass, and the
second by the overall electromagnetic energy momentum tensor
[recall that we take $O(e)=O(q)$].  We now argue that the
corrections to the area  formula actually appear only in the next
higher orders.

For suppose the area $A$ were indeed perturbed in linear
approximation to $O(\mu)$ and $O(e^2)$.  By spherical symmetry of
the background these corrections would not depend on the direction
along which the object was lowered.  If $n$ equal bodies were
lowered, each along a different radial direction, the perturbation
would be $n$ times larger by linearity of the approximation.  But
if enough bodies were disposed on a spherical shell concentric
with the black hole, the perturbation of the metric at the horizon
should tend to zero by Birkhoff's theorem\cite{MTW} that the
metric exterior to a spherical charged black hole is exactly RN if
the surroundings are spherically symmetric too.  We thus get a
contradiction unless we admit that the perturbations of $O(\mu)$
and $O(e^2)$ vanish in linear theory.  Any corrections to $A$ must
be of higher order, like $O(\mu^2)$, {\it etc.\/} Hence by
Eq.(\ref{area})
\begin{equation}
A =8\pi(2m^2 -q^2) + O(q^4/m^2) + O(\mu^2) +O(\mu e^2/m)
\label{area2}
\end{equation}
where we have included all possible second order terms of the
correct dimensions; $O(e^4/m^2)$ is subsummed in the $O(q^4/m^2)$
which is the remainder of the expansion of $A$ in powers of $q$.
Below we denote the above sort of corrections by the ellipsis
$\cdots$ .

The descent of the object, if sufficient slow, is known to be an
adiabatic process which causes no change in the horizon
area\cite{disturbing,Mayo}.  It follows that to the stated
accuracy, $m$ is unchanged in the course of the lowering process
itself because $q$ and $A$ are unchanged. When the object is
finally absorbed by the black hole, $m$ increases by ${\cal E}$
while $q$ is augmented by $e$; after the suspension machinery has
been withdrawn (if adiabatically done, this will cause no further
area increase\cite{disturbing,Mayo}), we get an unperturbed RN
black hole with mass $m+{\cal E}$ and charge $q+e$.

Calculating its horizon area from Eq.(\ref{area2}) and
substracting the area of what was at first an unperturbed RN black
hole of mass $m$ and charge $q$ (because $e$ was still distant),
we find the change
\begin{equation}
\Delta A = 8\pi(4m{\cal E}-2qe-e^2) +O({\cal E}^2) +\cdots
\end{equation}
Finally substitution of Eq.(\ref{energyfinal}) gives
\begin{equation}
\Delta A\geq (8\pi\mu R - 4\pi e^2) \left[1+O\left({\mu R\over
m^2}\right)\right] +\cdots \label{dA}
\end{equation}
Notice that the black hole parameters $m$ and $q$ have dropped out
from the dominant terms, in analogy with results for
uncharged\cite{BHentropy} or spinning\cite{Hod}  objects. The
minimum change in black hole  entropy, $\Delta A/4\hbar$
corresponding to the equality in (\ref{dA}), is thus a property of
the object itself. The entropy of the object cannot exceed  this
amount, lest the overall entropy of the world decrease upon the
object's assimilation.  We thus find the bound on the entropy of
an object of charge $e$, proper energy $E=\mu$ and radius $R$ to
coincide with Zaslavskii's proposal Eq.(\ref{zaslavskii}).

\section{Variant Employing a Grounded Black Hole}

In Sec.III the  black hole is electrically isolated so its charge
$q$ is fixed. One can consider a variant gedanken experiment
involving a black hole which is electrically grounded.  An
approximation of  this could be achieved by having a conducting
``wire'' connect the horizon to matter at large distances.  One
wonders whether a different, perhaps tighter, bound on entropy
would be obtained from a repetition of the above gedanken
experiment.  Here we show that the same bound arises despite the
differences in the energetics.

Eq.(\ref{potentialg}) in the Appendix gives the potential
$A_0^{(g)}(\rho,\theta)$ engendered by a charge $e$ located at
$\rho=a$ and $\theta=0$ in the vicinity of a grounded spherical
static black hole. This potential vanishes both at infinity and at
the horizon ($\rho=m/2$), and may be obtained by setting
\begin{equation}
q=-8ame(2a+m)^{-2} \label{charge}
\end{equation}
in the expression for $A_0(\rho,\theta)$, Eq.(\ref{potential1}).
Although now the hole's charge (aside from the image charge) is
controlled by the external charge's position,  the contribution of
$q$ to energy is still lumped with the black hole (see below).
Therefore, $q$ is still regarded as generating an external
potential for $e$. We may thus take over
Eq.(\ref{potentialenergy}) for the object's energy, but with $q$
replaced according to Eq.(\ref{charge}).

Expression (\ref{energy3}) for ${\cal E}$ is thereby replaced by
\begin{equation}
{\cal E} =\left({2\mu\ell-3e^2\over
8m}\right)\left[1+O\left({\ell\over m}\right)\right] +
O\left({e^4\over m^3}\right) \label{energy4}
\end{equation}
Neglecting the corrections we get in place of
Eq.(\ref{energyfinal}) the very different result
\begin{equation}
{\cal E}\geq {2\mu R-3e^2\over 8m} \label{E}
\end{equation}

Unlike the case discussed in Sec.III, here ${\cal E}$ is not the
exclusive contribution to the change in $m$.  When the object with
charge $e$ is very distant from it, the black hole is exactly
Schwarzschild [because $q$=0 by Eq.(\ref{charge})] with mass $m$
and horizon area $16\pi m^2$.  As the object approaches, charge
flows into  the hole through the wire, and $q$ varies according to
Eq.(\ref{charge}). Because the descent is slow, the change in the
hole is adiabatic; thus it should not cause growth of the
horizon's area\cite{disturbing,Mayo}.  We shall assume the current
flows reversibly (in the sense of Christodoulou's
transformations\cite{Christodoulou}), so that it does not cause a
change of area either.  According to Eq.(\ref{area2}), $m$ will
have to grow to compensate for the increase in $q$.  In the limit
$a\rightarrow m/2$, $q\rightarrow -e$ so that $m\rightarrow
m'\equiv \sqrt{m^2+e^2/2}\approx m+e^2/4m$.  When the object is
assimilated, its charge $e$ exactly neutralizes the hole's charge,
while its energy augments the hole's mass $m'$ to
$m''=m+e^2/4m+{\cal E}$.   The new horizon area is thus
$16\pi(m+e^2/4m+{\cal E})^2$.  Substituting ${\cal E}$ from
Eq.(\ref{E}), and substracting $16\pi m^2$ we obtain the overall
change in area
\begin{equation}
\Delta A\geq 8\pi\mu R\left[1+O\left({\mu R\over
m^2}\right)\right] -4\pi e^2 + \cdots
\end{equation}
This is the same as Eq.(\ref{dA}), so that from its minimum value
we reproduce the bound on entropy of a charged object,
Eq.(\ref{zaslavskii}).

How do we get the same area increase out of two different
expressions for the energy of the object, (\ref{energy3}) and
(\ref{energy4}) ?  The difference is compensated by a
complementary difference in the behavior of the electric potential
of the horizon. In the first case as $a\rightarrow m/2$,
$A_0\rightarrow -(e+q)/2m$.  In the second case
$A_0^{(g)}(m/2,\theta)=0$ identically.  Thus in the first case the
infall of the charge contributes a change in horizon area through
the change in black hole charge, while in the second it does not.

\section{The Optimal Bound}

For an object with spin $s$, charge $e$, maximal radius $R$ and
mass-energy $E=\mu$  we may, by comparing Eqs.(\ref{zaslavskii})
and (\ref{Hod2}), conjecture the tighter entropy bound
\begin{equation}
S\leq 2\pi {\sqrt{E^2R^2-s^2} - e^2/2\over\hbar} \label{hybrid}
\end{equation}
As a check we look at the case of slow rotation:
\begin{equation}
S\leq {2\pi R\over \hbar}\left[E-{s^2\over 2\mu R^2}-{e^2\over
2R}\right] +O(s^4) \label{approx2}
\end{equation}
Comparing with remarks made in Sec.II we see that here the maximum
possible Coulomb energy and ${\scriptstyle 2\over \scriptstyle 3}$
of the maximum possible rotational energy are deducted from the
total energy, with the remainder taken as the $E$ in the original
entropy bound (\ref{firstbound}).  Obviously, for a
nonrelativistically rotating ordinary object, bound
(\ref{approx2}) is correct, and on the liberal side.

This correspondence argument does not prove the correctness of
bound (\ref{hybrid}); that bound is not the unique progenitor of
the nonrelativistic form (\ref{approx2}).  In addition, one could
argue that there seems to be something missing from
Eq.(\ref{hybrid}).  A spinning charged object has a magnetic
dipole moment proportional of $O(es/\mu)$ which generates a
magnetic field, and thus contributes to the electromagnetic
energy.  We see no such contribution reflected in bound
(\ref{approx2}). However, it must be recalled that magnetic dipole
energy is of higher order in $c^{-1}$ than Coulomb energy.  If we
care about this higher order, we should continue the Taylor
expansion of the root in (\ref{hybrid}) to $O(s^4)$ which is of
the same order.  However, we have just mentioned that bound
(\ref{approx2}) understates the amount of rotational energy in the
system by a substantial factor already at $O(s^2)$.  It is thus
pointless to go to higher order in rotational or electromagnetic
energy.  Bound (\ref{approx2}) is not strict, but liberal, and so
is bound (\ref{hybrid}).  Thus at present we find no reason to
cast doubt on the general bound (\ref{hybrid}).

A more positive point for bound (\ref{hybrid}) is the fact that
any KN black hole (mass $m$, charge $q$ and angular momentum $j$)
saturates it.  The horizon area of such black hole is\cite{MTW}
\begin{equation}
A = 4\pi(r_+^2+j^2/m^2);\qquad r_+\equiv m+(m^2-j^2/m^2-q^2)^{1/2}
\label{AKN}
\end{equation}
Substituting $r_+$, squaring as required, and cancelling terms
gives
\begin{equation}
A = 2\pi(4mr_+-q^2) =  2\pi[4m(r_+^2+j^2/m^2-j^2/m^2)^{1/2}-q^2]
\label{next}
\end{equation}
In light of Eq.(\ref{AKN}) it is reasonable interpret
$(r_+^2+j^2/m^2)^{1/2}$ as the radius $R$ of the hole.
Incorporating this in the last equation and dividing by $4\hbar$
gives for the black hole entropy
\begin{equation}
S_{\rm BH} ={2\pi\over \hbar}[(m^2R^2-j^2)^{1/2}-q^2/2]
\label{entropy}
\end{equation}
If we identify $m\leftrightarrow E$, $q\leftrightarrow e$ and
$j\leftrightarrow s$, this is exactly the upper bound of
Eq.(\ref{hybrid}).  Hence the KN black hole saturates the proposed
entropy bound.  This property would be lost if modifications were
made to the bound. Hence we adopt it in the form given.  Study of
the role of spin--curvature effects in the discussion in Sec.III
is in progress in order to provide a more direct argument for the
full bound (\ref{hybrid}).

Parenthetically  we should mention another way to look at the
saturation. Suppose we had some means to slowly lower a small KN
black hole with mass $\mu$, charge $e$ and angular momentum $s$
into a much  larger KN black hole with corresponding parameters
$m$, $q$ and $j$.  Then the black holes would merge reversibly,
{\it i.e.,\/} with no overall growth in horizon area.  This is
obvious because {\it if\/} bound (\ref{hybrid}) can be derived by
the arguments expounded in Secs.III and IV, then the overall
growth in area of the big black hole must correspond precisely to
the equality case in bound (\ref{hybrid}) (for parameters $\mu$,
$e$ and $s$) multiplied by $4\hbar$.  But this just says that the
big horizon expands by precisely the area of the small horizon, so
that the merged horizon has area equal to the sum of the two
original ones.

Bound (\ref{hybrid}) is readily generalized to include magnetic
monopole charge $g$.  Duality of electromagnetism leaves little
doubt that one should just replace $e^2\rightarrow e^2+g^2$.  The
deeper question arises, can one give generic bounds on entropy
which are tighter than (\ref{hybrid}) by virtue of the object
possessing some conserved ``quantum number'' apart from $q$, $g$
or $s$ ?  A case in point would be a tighter bound for an object
with definite and known baryon number.  We now marshal evidence in
support of the conjecture that bound (\ref{hybrid}), with the
extension to magnetic monopole, cannot be bettered generically.
By ``generically'' we mean without knowledge of details about the
object's structure and dynamics.  When these are known it is
possible to compute by means of statistical mechanics bounds on
the entropy which can be small compared to bound
(\ref{firstbound}), for example\cite{Numer}.  But if we use no
such information, we must go back to the black hole derivation of
the entropy bounds, and it is for this situation that we
conjecture that bound (\ref{hybrid}) cannot be bettered.

The ``no hair'' conjecture is central to our argument.  A large
amount of work has certified that a stationary black hole can have
just a few parameters. The incontestable ones are mass, charge,
magnetic monopole and angular momentum.  Skyrmion number is an
extra possibility\cite{skyrmion}, but one whose physical
significance is unclear\cite{Brazil}.  Other candidates, such as
color\cite{color}, scalar charge\cite{BBM} and massive Yang--Mills
charge\cite{Greene}, are associated with unstable black
holes\cite{Brodbeck}. The sort of arguments we have given in
Secs.III and IV make sense only if the black hole is stable to
outside perturbations.  Hence we focus on the KN black holes with
parameters $m$, $q$, $g$ and $j$.

Suppose we add to such a black hole an object carrying an extra
additive conserved quantity $b$.  If $b$ is a ``global'' quantity,
such as baryon or lepton numbers are thought to be, it generates
no field of its own.  The energy--momentum tensor originating in
the object is thus unaffected by $b$ (put another way, effects of
$b$ can be absorbed in the mass).  Hence, even if perturbations of
the metric are taken into account, $b$ cannot directly perturb the
horizon area formula (\ref{AKN}), and so $m$ is unaffected by slow
(adiabatic) lowering of the object.   Further, absence of $b$ from
the list of black hole parameters means the black hole has no
chemical potential conjugate to $b$.  Thus when the object finally
enters the hole, it cannot change the horizon area except through
the change in $m$, which is ${\cal E}$.  But ${\cal E}$ gets {\it
no\/} contribution specific to $b$ since the latter does not
generate a field that could polarize the hole, {\it c.f.\/}
Eq.(\ref{energy}).  Therefore, the change in horizon area is
$b$--independent.   But then the bound that can be set on the
entropy by the argument of Sec.III is also independent of $b$: the
new quantity does not allow tightening of the entropy bound.

Much the same conclusion can be reached if $b$ generates a short
range field, schematically denoted by ${\cal B}$.  For example,
$b$ could be weak hypercharge, a source of the short range,
$Z$--boson mediated, weak force. Although there is now a
contribution to the energy--momentum tensor from ${\cal B}$, it is
localized around the object, and thus can be lumped into its usual
energy--momentum tensor.  No novel perturbation to the metric
arises from this.  Hence, $b$ cannot directly perturb the horizon
area formula (\ref{AKN}), and so $m$ is unaffected by slow
lowering of the object.  Furthermore, no novel potential term is
contributed to ${\cal E}$ by ${\cal B}$ unless the particle is
already next to the horizon; otherwise the short range field
${\cal B}$ does not reach down to the horizon and cannot polarize
it.  Hence, in this case also, the change in horizon area turns
out to be  $b$--independent, and $b$ cannot appear in a generic
entropy bound.

The  third and last case is when $b$ is the source of a
long--range field, again denoted ${\cal B}$. The range may be
finite if large compared to typical object size.  Now the area
formula may differ from Eq.(\ref{AKN}) by terms depending on $b$
because of the perturbation that ${\cal B}$'s energy momentum
tensor exerts on the metric.  Unless ${\cal B}$ is a gauge field
which (unlike the $Z$ and $W$ boson fields of Weinberg--Salam
theory) remains massless in the classical (or low energy) limit,
we cannot rule out such dependence, as we did in Sec.III.  This is
because Birkhoff--type theorems exist only for massless vector
fields, and from our point of view the electromagnetic field is
the only one such, and has already been accounted for in Secs.III
and IV.  Thus, while the area stays constant during the descent as
required by the adiabatic theorem, $m$ may change by a quantity of
$O(b^2)$ as the object descends.  The sign of this quantity is
unclear without a specific model.

In addition, ${\cal E}$ is most likely to have a term of $O(b^2)$
for the same reasons as in Eq.(\ref{energy3}) (by ``no hair''
there is now no analog of $q$).  It may even be that this term is
also positive here, yet it does not follow that the effect of $b$
is to suppress the growth of area, as it did in Sec.III, because
of the correction of indefinite sign to the area formula. Thus
without calculating linear corrections to the metric, one cannot
settle the question of whether the change in area is incremented
or depressed by $b$'s presence.   However, we have as yet
uncovered {\it no\/} clear evidence that an improved bound
(\ref{hybrid}) will result for a long--range field which is not a
massless vector field.  The conjecture that bound (\ref{hybrid})
is the tightest {\it generic\/} bound on entropy thus seems
reasonable.

{\bf ACKNOWLEDGMENTS} We thank Shahar Hod for discussions, and
Bernard Linet for pointing out a crucial omission.  This research
is supported by a grant from the Israel Science Foundation,
established by the Israel Academy of Sciences and Humanities.

\appendix
\section{Copson--Linet solution for charge in black hole background}

Here we determine $A_0$ resulting from a charge $e$ in the
Schwarzschild background (\ref{metric}).  Using the conventions of
Misner, Thorne and Wheeler\cite{MTW} we write the electromagnetic
field as $F_{\alpha\beta}=A_{\beta ,\alpha} - A_{\alpha ,\beta}$.
We express the Maxwell equations $F^{\alpha\beta}{}_{;\beta} =
4\pi j^\alpha$ for the axisymmetric stationary field of a test
point charge $e$ situated at $\{\rho,\theta\}=\{a,0\}$ as
\begin{equation}
\left[{\rho^2\,A_{0,\rho}\over (1-{m\over 2\rho}) (1+{m\over
2\rho})^2}\right]_{,\rho} +{A_{0,\theta\theta}\over (1-{m\over
2\rho})(1+{m\over 2\rho})^2} =4\pi e
\delta(\rho-a)\delta(\theta)\delta(\phi) \label{Laplace}
\end{equation}
The source term takes on the indicated form because the relevant
3--D Dirac delta function has the form
$(-g)^{-1/2}\delta(\rho-a)\delta(\theta)\delta(\phi)$.

A convenient solution of this equation was found by
Copson\cite{Copson} long ago:
\begin{equation}
A_0^{\rm (C)}(\rho,\theta)=-e{\chi+{m^2\over 4a^2\chi}\over
(1+{m\over 2\rho})^2\rho (1+{m\over 2a})^2} \label{Copson}
\end{equation}
where
\begin{equation}
\chi(\rho,\theta)\equiv \left[{\rho^2+(m^2/4 a)^2 -2 (m^2/4a)\rho
\cos\theta\over \rho^2+a^2-2a\rho\cos\theta}\right]^{1/2}
\label{mu}
\end{equation}
In $A_0^{\rm (C)}$ there appears in denominators not only the
Euclidean distance $(\rho^2+a^2-2a\rho\cos\theta)^{1/2}$ between
the field point $\{\rho,\theta\}$ and the charge's position, but
also the distance of the former from the point $\{m^2/4a,0\}$,
which is the appropriate location for the image charge in the
solution of Laplace's equation for a charge near a conducting
sphere of radius $m/2$ by the method of images.  This is
consistent with the expectation that the black hole is polarized
by influence of the charge $e$.

As noted by Linet\cite{Linet}, the coefficient of the $1/\rho$
term in the asymptotic form of this potential indicates that a
total of charge $e-e(m/a)(1+m/2a)^{-2}$ resides in the spacetime.
The charge $e$ being the only source outside the black hole, one
must perforce admit that the black hole bears charge $\tilde
e=-e(m/a)(1+m/2a)^{-2}$.  Of course such charge must modify the
metric, as does the exterior charge $e$.  But such perturbations
will be of order $e^2$ and may be ignored in computing $A_0^{\rm
(C)}$ correct to $O(e^2)$.  Linet proposes that a more relevant
solution to the problem is to be  had by adding to $A_0^{\rm (C)}$
a monopole field with charge $-\tilde e$.  We shall push this a
little farther and add to $A_0^{\rm (C)}$ the monopole potential
appropriate to charge $q-\tilde e$.  Since the spherically
symmetric, everywhere regular, solution of Eq.(\ref{Laplace}) is
$\rho^{-1}(1+{m\over 2\rho})^{-2}$, we must write
\begin{equation}
A_0(\rho,\theta)=A_0^{\rm (C)}(\rho,\theta) - {q+4{ema\over
(2a+m)^2}\over \rho(1+{m\over 2\rho})^2} \label{potential1}
\end{equation}
The charge in the spacetime is now $q+e$, with $q$ in the black
hole, as it should.  So long as $q$ is of order $e$ and this last
is small on the scale of $m$, we do not have to correct the metric
or Eq.(\ref{Laplace}) to get $A_0$ correct to $O(e^2)$.

Potential (\ref{potential1}) has the constant value
$-q/2m-4ae(2a+m)^{-2}$ on the horizon ($\rho=m/2$).  One can ask
the question, what would be the potential if the charge $e$ were
to coexist with a black hole which is grounded.  In practice this
could be achieved by having a conductor connect the horizon to
matter at large distances.  The desired solution of
Eq.(\ref{Laplace}) is now one satisfying
$A_0(m/2,\theta)=A_0(\infty,0)=0$.  It is easily checked that the
desired potential can be gotten from $A_0$ in
Eq.(\ref{potential1}) by setting $q\rightarrow -8ame(2a+m)^{-2}$.
We denote it by $A_0^{(g)}$ (``g'' for ``grounded''):
\begin{equation}
A_0^{(g)}(\rho,\theta)=A_0^{\rm (C)}(\rho,\theta) + 4{{ema\over
(2a+m)^2}\over \rho(1+{m\over 2\rho})^2} \label{potentialg}
\end{equation}
Of course, in the present case the charge on the black hole varies
with $a$; this is because as the charge $e$ draws near the black
hole, opposite charges are drawn into the hole through the
conductor.

\end{document}